# Specificity of Trypsin and Chymotrypsin: Loop Motion Controlled Dynamic Correlation as a Determinant

Wenzhe Ma[1,3], Chao Tang[1,2], and Luhua Lai[1,3]*

[1]Center for Theoretical Biology, Peking University, Beijing 100871, China

[2]California Institute for Quantitative Biomedical Research

Departments of Biopharmaceutical Sciences and Biochemistry and Biophysics

UCSF Box 2540

University of California San Francisco

San Francisco, CA    94143-2540

[3]State Key Laboratory for Structural Chemistry of Stable and Unstable Species,

College of Chemistry, Peking University, Beijing 100871, China

* Corresponding author: Email: lhlai@pku.edu.cn

Running title:    Loop Motion and Substrate Specificity

# ABSTRACT

Trypsin and chymotrypsin are both serine proteases with high sequence and structural similarities, but with different substrate specificity. Previous experiments have demonstrated the critical role of the two loops outside the binding pocket in controlling the specificity of the two enzymes(1). To understand the mechanism of such a control of specificity by distant loops, we have used the Gaussian Network Model to study the dynamic properties of trypsin and chymotrypsin and the roles played by the two loops. A clustering method was introduced to analyze the correlated motions of residues. We have found that trypsin and chymotrypsin have distinct dynamic signatures in the two loop regions which are in turn highly correlated with motions of certain residues in the binding pockets. Interestingly, replacing the two loops of trypsin with those of chymotrypsin changes the motion style of trypsin to chymotrypsin-like, whereas the same experimental replacement was shown necessary to make trypsin have chymotrypsin's enzyme specificity and activity(1). These results suggest that the cooperative motions of the two loops and the substrate-binding sites contribute to the activity and substrate specificity of trypsin and chymotrypsin.

**[INTRODUCTION]**

Serine proteases include a large class of enzymes. They provide much information on enzyme catalysis(2). Catalytic triad and oxyanion hole are important for enzyme activity of this category(3,4). These enzymes bypass the obstacles of breaking a peptide bond by properly positioning the catalytic triad(5), passing proton through them and forming catalytic intermediate(6,7)，and stabilizing the tetrahedral intermediate with the oxyanion hole by electrostatic complementarities(8). Specificity is another aspect of enzyme catalysis. It is closely related to the enzyme-substrate interaction. From a mechanistic point of view, specificity is largely determined by the binding and the acylation step(2). Residues such as 189, 216 and 226 are important specificity determinants in these enzymes(9,10).

Hedstrom gave a thorough description in her recent review(2) about serine protease. Despite a long time study, many aspects of this class of enzymes are still unclear. It is even not clear what the rate-limiting step in such proteases is. For poor amide substrates, acylation step seems to be rate-limiting(11), whereas there is evidence that in serine protease like Kex2, deacylation step is rate-limiting(12).

Trypsin and chymotrypsin are both serine proteases. The two enzymes have high sequence identity(13) and their tertiary structures are very similar (Fig. 1A). In the chymotrypsin index, His57, Asp102 and Ser195 form the catalytic triad, residues 189-195, 214-220 and 225-228 form the primary substrate-binding pocket called S1 binding pocket. Residues 185-188 and 221-224 form two loops near the S1 pocket, called L1 and L2, respectively (Fig. 1B). Catalytic mechanisms of these two proteases are similar, but their substrate specificities are different. Trypsin favors basic residues like lysine and arginine; chymotrypsin favors aromatic residues like phenylalanine, tyrosine and tryptophan(14). The S1 binding pocket in trypsin and chymotrypsin are almost identical in primary sequences and backbone tertiary structures (Fig. 1). An important difference is that residue 189 is a negatively charged Asp in trypsin and a polar Ser in chymotrypsin. This residue lies at the bottom of the S1 binding pocket and determines different S1 pocket chemical properties. This difference was once used to explain the different specificity of trypsin and chymotrypsin(15). But the mechanism is not that simple. Mutation of Asp189 in trypsin (D189S) did not change the substrate specificity from trypsin-like to chymotrypsin-like(1,16,17), instead the enzyme just lost its activity. And mutation of S189D in chymotrypsin did not convert its specificity into that of trypsin, either(18). Comparison between trypsin and trypsin mutant(D189S) shows little structural change in the S1 binding pocket(19). Rutter et al. showed that the S1 binding pocket only determines the specificity of ester hydrolysis, whereas specific amide hydrolysis requires both the proper S1 binding site and more distal interactions such as loops beside the substrate-binding pocket(1). When the two loops L1 and L2 of trypsin were replaced by those of chymotrypsin in addition to the D189S mutation, the new protein shows an increase of chymotrypsin activity to about 1000 fold against the D189S mutant(1). A site mutation not in contact with the substrate (Y172W) was found to improve the chymotrypsin-like activity of the hybrid protein by 20-50 fold(20). Gly216 was also found to be a specificity determinant(21). The backbone conformation of Gly216 differs between trypsin and chymotrypsin; but the hybrid enzyme adopts a chymotrypsin-like conformation(10,16,21,22). These experiments imply that in addition to the S1 substrate-binding pocket, loop regions of trypsin and chymotrypsin have significant effect on enzyme activity and substrate specificity.

Several explanations about the experiments on the specificity change have been proposed. An obvious one is from structure. The substitution of D189S deforms the S1 site and the activation domain(2,16,23). Mutations on L1 and L2 loops, and on Y172W may help to stabilize the S1 site(2,10). Though the specificity of chymotrypsin-like serine protease is usually categorized in terms of the P1-S1 interaction, a crucial feature of these proteases is that substrate occupancy of the S1 binding site alone confers only modest specificity(2). L1, L2 substitutions affect the conformation of Gly216, which is an important residue to bind the P3 residue. Crystal structures show that the conformation of Gly216 becomes chymotrypsin-like in the hybrid protein and help to orientate the scissile bond in the enzyme complex structure(21). Question remains as how the L1, L2 substitutions change the conformation of Gly216.

The above argument is from the static point of view. The other possibility is that the dynamical properties of the enzymes play an important role in the catalytic process. It is known in many cases that structure flexibility is closely related and crucial to the enzyme activity(24-27). A study of α-lytic protease has shown that plasticity of the substrate binding pocket affects specificity of the enzyme(28). Studies on lipase showed that enzyme catalysis, substrate binding, and substrate releasing correspond to different type of motion styles(29). Enzyme loop regions have been shown to be important in catalysis(1,30-35). For trypsin-chymotrypsin system, it is possible that certain modes of motion are essential for chymotrypsin catalysis, which can be influenced by the L1 and L2 loops. If only trypsin S1 pocket is changed into chymotrypsin like, it is not sufficient to change the specificity; but when L1 and L2 are also changed, global dynamics of the protein may change to benefit the catalysis.

In the present study, we have used the Gaussian Network Model (GNM)(36) and a clustering method to analyze the dynamic properties of trypsin and chymotrypsin. We find that the two enzymes have certain key differences in their dynamic motion. In particular, they differ in ways that the motion of the S1 binding pocket correlates with that of the loops L1 and L2, and with the nearby regions. When the two loops in trypsin are replaced to those of chymotrypsin, the hybrid enzyme vibrates in a similar way as chymotrypsin in some key parts. Taken together with experimental findings(1,21,37), our results suggest that the concerted motions of loop regions with the S1 binding pocket and the correlations between different binding sites can be important for the enzyme specificity.

**[MATERIALS AND METHODS]**

1. Gaussian Network Model

Gaussian Network Model (GNM) is a simplified model for normal mode analysis of proteins(36), in which a protein is converted into nodes connected by springs. All the nodes are identical and each of them represents a single residue. We use $C_\alpha$ atoms as the nodes in this study. All the nodes within a given distance $r_c$ have interactions with each other. The connection here is simplified as harmonic force, with the same force constant. The distance of $r_c$ is defined as 7Å. This value comes from the results of statistical analysis(38,39). All other atomic and structural details are ignored. This coarse-grained model was successfully used to reproduce the B-factors in X-ray diffraction experiment(40) and NMR experiment(41), to find kinetically hot residues(42), and to study relationships between slow vibration modes and the protein function(36,43,44).

The dynamics of the protein is controlled by the connectivity (or Kirchhoff) matrix $\Gamma$. Elements of $\Gamma$ are defined as(40):

$$\Gamma_{ij} = \begin{cases} -1 & i \neq j \quad and \quad r_{ij} \leq r_C \\ 0 & i \neq j \quad and \quad r_{ij} > r_C \\ -\sum_{i, i \neq j} \Gamma_{ij} & i = j \end{cases} \quad (1)$$

where $r_{ij}$ is the distance between the $C_\alpha$ atoms of residues $i$ and $j$. $\Gamma_{ij}$ =-1 ($i \neq j$) means that residue $i$ and $j$ have a spring connection, that is, they have interaction, and $\Gamma_{ij}$=0 means that there is no connection. The potential of the system is $V = \frac{\gamma}{2}(\Delta R)^T \Gamma (\Delta R)$. $\Delta R$ is a vector, with $\Delta R_i$ denoting the displacement of the $i$th residue from its equilibrium position. In GNM, each residue has only one degree of freedom; x, y and z directions are treated the same (they decouple). We should note that $\Delta R_i$, which can be either positive or negative, has certain directional information. The correlation between $\Delta R_i$ and $\Delta R_j$ reflects whether the two residues move in the same way or not. The correlation is positive if they move in the same direction and is negative if they move in the opposite direction. The equilibrium correlations between the fluctuations $\Delta R_i$ and $\Delta R_j$ of residues i and j are given by(40,45,46):

$$<\Delta R_i \cdot \Delta R_j> = \frac{1}{Z_N}\int(\Delta R_i \cdot \Delta R_j)\exp(-\frac{V}{k_BT})d\{\Delta R\} = (k_BT/\gamma)[\Gamma^{-1}]_{ij} \qquad (2)$$

where $Z_N$ is the partition function of this system:

$$Z_N = \int \exp(-\frac{V}{k_BT})d\{\Delta R\}. \qquad (3)$$

From $<\Delta R_i^2>$, we can get Debye-Waller or temperature factors(47):

$$B_i = 8\pi^2 <\Delta R_i^2>/3 \qquad (4)$$

This is what we use to compare with the experimental temperature factor. In GNM, the correlation is normalized as:

$$C_{ij} = \frac{<\Delta R_i \cdot \Delta R_j>}{[<\Delta R_i^2>\cdot<\Delta R_j^2>]^{\frac{1}{2}}} = \frac{[\Gamma^{-1}]_{ij}}{([\Gamma^{-1}]_{ii}[\Gamma^{-1}]_{jj})^{\frac{1}{2}}} \qquad (5)$$

where $[\Gamma^{-1}]_{ij} = \sum_{k=2}^{N}\frac{u_{ik}\cdot u_{jk}}{\lambda_k}$, with $u_{ik}$ being the *i*th entry of the *k*th eigenvector, $\lambda_k$ being the *k*th eigenvalue. Because in GNM, the first mode is simply the translation, we sum over the remaining N-1 modes. Correlation value ranges between –1 and 1, the higher the absolute value; the more the two residues are correlated. Using a modified GNM, Micheletti et al. have shown that the correlations from molecular dynamics simulation and their modified GNM are similar. And the simplified model was successfully used to identify important correlated motions related to HIV-1 protease catalysis(48).

2. Correlation analysis

Once we have the correlation matrix $C_{ij}$, one way to use the matrix is to plot the matrix on a 2D map, just like Figure 2. This plot have been used in several studies(44,49-52). However, this map can only make clear correlations within and between big cliques of consecutive residues. Here we analyze the data in an alternative way. We change the correlation map into a distance map, and use clustering methods to analyze it. Similar procedures have been widely applied in genetic evolutionary analysis(53,54).

In our analysis we define $d_{ij}=1-|\widetilde{C}_{ij}|$ as distance (ranging 0 to 1). $d_{ij}$ is the element of distance matrix $D$. The definition of $\widetilde{C}_{ij}$ is similar to $C_{ij}$ in Eq. 5, but the correlation between residue $i$ and $j$ is calculated only with a predefined number of modes. We want to study the relationship between the L1, L2 loops and the rest of the protein. As low frequency modes often correspond to functional motions which include distant residues and high frequency modes correspond to localized motions(55), only low frequency modes are used here to improve the "signal-to-noise" ratio. Specifically, we use the formula:

$$\widetilde{C}_{ij} = \frac{[\Phi^{-1}]_{ij}}{([\Phi^{-1}]_{ii}[\Phi^{-1}]_{jj})^{\frac{1}{2}}} \tag{6}$$

$$[\Phi^{-1}]_{ij} = \sum_{k=2}^{m} \frac{u_{ik} \cdot u_{jk}}{\lambda_k}$$

where $m$ is the mode number of the highest frequency mode used in the calculation. We use $m=40$ in our calculations because we can see from Figure 3A that the fluctuation amplitude changes little after mode 40. More modes were also tried and gave similar results with weaker signals. As both positive and negative values indicate correlations and only the absolute values are meaningful, we use a modified distance definition $d_{ij} = 1 - |\widetilde{C}_{ij}|$ rather than conventional distance definition: $d_{ij} = 1 - \widetilde{C}_{ij}$. After we get the distance matrix *D*, we can use program KITSCH in PHYLIP (http://evolution.genetics.washington.edu/phylip.html) to analyze the clustering properties.

The crystal structure coordinates for bovine α-chymotrypsin(56) (PDB code: 4CHA), bovine trypsin(Chamorro Gavilanes,J.A., J.A.Cuesta-Seijo, and S.Garcia-Granda. 2005. Pancratic Bovine Trypsin Native and Inhibited with Benzamidine from Synchotron Data. *To be Published*, PDB code: 1S0Q) are used in this study.

**[RESULTS AND DISCUSSIONS]**

1. Correlation map

The correlation map $C_{ij}$ (Eq. 5) of chymotrypsin is shown in Figure 2. A number of features are evident. First, there are two highly correlated small squares at the diagonal around residues 160 and 235, respectively; these squares correspond to the two α helices in chymotrypsin. The motions of the residues within each helix are highly correlated, implying that the alpha helix is a compact and relatively independent structure motif with its own coherent motions(57). Second, there are several short lines of high correlation across and perpendicular to the diagonal (40-50, 57-62, 70-80, 100-110, 128-138, 142-152, 170-180, 192-197, 210-220). These correspond to beta sheets in the protein structure. Note that the correlation map shows certain information about the secondary structures though the model itself does not contain secondary structure information explicitly. Thirdly, there are two large weakly correlated regions in the bottom-left (10-120) and top-right (125-155, 175-220) of the map. These two regions correspond to the two β barrels of chymotrypsin. No other large correlated movement can be seen from the map. In chymotrypsin, the smallest correlation is about -0.1 and in other systems like HIV reverse transcriptase(49) the correlation could be more negative. This negative correlation is related to the specific structure and functional motion of the proteins. HIV reverse transcriptase is composed of many domains, motion between domains is functionally important. However, besides structural reasons, the correlations in the paper of Bahar et al. were enhanced because they only used the first 4 modes. If more modes are used, there will be more local fluctuations which do not contribute to the domain-domain correlation and due to the normalization with more modes their correlation values will be smaller. It is important to note that mode number have different effect on the max value of positive and negative correlations. The positive correlations exist among nearby residues, they often have the same motion style in most modes (especially the self-correlation), so that the mode number will not affect the positive correlation much. But the negative correlation can not exist among nearby residues; they will be affected by the mode number. Trypsin and chymotrypsin are relatively "stiff" enzymes; they do not have very long loops and also we use all the modes here so there are no big negative correlations.

2. Clustering analysis

After clustering the distance matrix of the pair-wise correlations (Eq. 6), we obtain a tree map in which highly correlated residues cluster together (Fig. 3B). These clusters provide dynamical information of the protein structure in addition to the traditional static view of protein domains, which may be functionally relevant. In Figure 3C several clusters are shown on the three-dimensional structure of chymotrypsin. Different clusters are painted with different colors. We can see that both L1 and L2 are located in the purple region together with residues in the S1 pocket Ser189, Ser214, Trp215, Gly216 and Gly226. Ser189, Gly216 and Gly226 help define a deep hydrophobic pocket with other residues in chymotrypsin. Residues 214-216 have interactions with the P1-P3 residues of a peptide substrate. Next we focus on the local tree branch near the L1-L2 loop of chymotrypsin in Figure 4A. In this figure, residues in the L1-L2 loops (shown as solid circles) and some residues in the substrate binding pocket (solid triangles) are clustered together, so they move coherently. For trypsin, we also run this procedure and get a similar clustering map, which is shown in Figure 4B. Residues in L1-L2 loops and several residues in the S1 binding pocket also cluster together, but the topology of the tree has changed. One obvious change is that in chymotrypsin, residues on the lid of the S1 pocket (217, 218 and 219) correlate with L1 and L2 loops stronger than that in trypsin. We have known from experiments that loop replacement helps to change trypsin specificity to chymotrypsin specificity(1). Here we do the same experiment in *silico* by replacing the loops of trypsin with the loops of chymotrypsin. L1 structure of this hybrid protein is not known, but the backbones of the L2 loop in hybrid protein and chymotrypsin are similar(21). We assume that the configurations of the L1 loop do not change much from chymotrypsin to the hybrid protein. Because GNM is a coarse-grained method, it is reasonable to replace these region directly after structure superposition (we changed L1-L2 loops and 217-219). Figure 4C shows local tree map for the hybrid protein by using the first 40 modes in the calculation. We see that L1-L2 move coherently with several residues in the S1 binding pocket, just like in chymotrypsin. In particular, the lid of the pocket (217-219) clusters with L1, L2 loops closely. In the hybrid protein, we get similar dynamic performance as chymotrypsin. It is noteworthy that residue 138, 184-186, 188-189, 192, 217, 221-224 in trypsin were mutated(1) in the experiment (Fig. 1B). Most of them can be found in one big branch of the tree - at least 13 in 15 of these resides appear together in the big branch for trypsin (Fig. 4A), 9 in 15 for the hybrid protein (Fig. 4C). This may imply that these residues cooperate with each other to fulfill their function.

As we already knew that L1 and L2 correlate strongly with the S1 binding pocket in particular with several important binding sites like Gly216 and Gly226, we further analyze the differences of key residue correlations to see what happens when the loops are substituted. Specifically, we define a parameter:

$$S_{ij}(similarity) = \left| \widetilde{C}_{ij;the\ hybrid\ protein} - \widetilde{C}_{ij;tryp\sin} \right| - \left| \widetilde{C}_{ij;the\ hybrid\ protein} - \widetilde{C}_{ij;chymotryp\sin} \right| \quad (7)$$

where *i* and *j* are residue indices. If $S_{ij}$ is bigger than 0, it means that the correlation value of the hybrid protein is closer to that of chymotrypsin than trypsin, and vise versa. In the analysis, we only use important residues for binding and catalysis (57, 102, 195: catalytic triad; 16,193-195: oxyanion hole; 189-192, 214-216, 224-228: S1 site; 57, 215, 99: S2 site; 172: important for activity; 142-143,151: S2' site; 41-45,

55-59: S1', S3' sites)(2).  From the result (Fig. 5) we can see that most of the S value is smaller than 0, that is that those correlations are trypsin like, which is natural because most of the residues in the hybrid protein are intact.  Meanwhile some of the correlations are chymotrypsin like.  The most important ones of those residue pairs are denoted in Fig. 5.  Among these residues, residue 189 is in the bottom of S1 pocket and is the most important residue in the pocket.  Residue 216 forms two hydrogen bonds with the ligand and was considered to be a specificity determinant in trypsin-chymotrypsin(21).  Residue 226 is used to create a negatively charged S1 site that accounts for trypsin's specificity(58).  Residue 172 substitution can improve the activity of the hybrid protein by 50 folds(20).  The correlation of these important residues become chymotrypsin like after the loops were substituted; this implies that these residues may function in a cooperative way to determine the specificity.  We should note that most of the residues interact with residues 224 and 225.  Residue 224 is in the S5-S6 sites and residue 225 is in the S1 site.  It implies that loop substitution changed the relationship between S1/S5-S6 sites and the other binding sites.  This is in good agreement with the experiment that longer substrates have clearer specificity tendency (20) because the correlation effect becomes clear in longer substrates.  We want to declare that the "perturbation" of loops can pick out important residues that have been proved by experiments.  Also there are clear correlations of residue pairs such as 99-57 that are trypsin like.  Residue 99 is one of the residues in the S2 binding site and His57 functions in the catalytic triad to transfer proton.  The trypsin-like correlations as this one are the possible reason that the activity and specificity of the hybrid protein is still not fully recovered.

3. Mode analysis:

The clustering analysis shows that residues in the L1, L2 loop and the lid (residues 217-219) correlate differently in the two enzymes. Note that this is the part that have been changed in the experiments(1). We further analyze the most correlated residue-pairs to find out more information from the correlations. We define the total correlation of loop region as:

$$TC = \sum_{i<j} \widetilde{C}_{ij} \qquad (8)$$

where $i, j \in$ residues in the loop region and the pocket lid. For chymotrypsin, there are 4 residues in loop L1, 4 residues in loop L2, and 3 on the pocket lid, so that there are 11*10/2=55 residue pairs. $i$ and $j$ are residue indices among these residues. Every eigenmode should have a definite contribution to the total correlation, either positive or negative. This contribution is represented in the form:

$$TC_k = \sum_{i<j}(\widetilde{C}_{ij})_k = \sum_{i<j} \frac{\frac{u_{ik} \cdot u_{jk}}{\lambda_k}}{([\Phi^{-1}]_{ii}[\Phi^{-1}]_{jj})^{\frac{1}{2}}} \qquad (9)$$

It is the contribution of the $k$th eigenmode to the total correlation. The symbols in Eq. 9 are the same to those in Eq. 6. We normalize these contributions by dividing them by a constant $c = (\sum_{k \in \text{modes used in calculation}} TC_k^2)^{1/2}$. The normalized contributions from each mode are shown in Figure 6. We can see that low frequency modes contribute most to the loop region correlation and modes with their index bigger than 15 have almost no contributions. The fact that low frequency motions correlate with protein function has been proposed and supported by many studies(59,60,61,55,62). Our work here provides further evidence that low frequency fluctuations can be closely related to the protein's function. From Figure 6A, we see that several modes are particularly important (Y-axis value bigger than 0.15). For trypsin, they are modes 3 and 9. For chymotrypsin, they are modes 3, 4, 5, 6 and 11. For the hybrid protein, modes 3, 4, 5, 9 and 10 are the most important. There is a clear trend that in the hybrid protein, more low frequency modes participate in the correlated motion of the loop regions, just like that in chymotrypsin.

To see how the loop motion influences the dynamics of the whole protein, we use only the most important modes for the loop motion listed above for the three proteins to calculate the residue fluctuations of the entire protein (Fig. 6B). It is clear that after the loop substitution, fluctuations of the hybrid protein become similar to chymotrypsin, although it still has a trypsin backbone. The most obvious example comes from residues 85-105, which are not in the two loop regions, where in chymotrypsin there is big fluctuation and in trypsin the fluctuation is small. When the loops of trypsin are changed into that of chymotrypsin's, a peak appears in this region, showing that these residues have collective motions with the loops of chymotrypsin that are being placed in the hybrid protein. It is notable that one of the catalytic residues, Asp102, and the essential residues Leu99 for the S2, S4 substrate binding sites are in this region. The different dynamical relationships between the two loops and these sites in trypsin and chymotrypsin may have functional implications on the two different enzymes.

Figure 6C, 6D show some of the important modes we have identified. Mode 3 shown in Figure 6C is a common mode that has big contribution in all the proteins. Mode 11 in chymotrypsin, mode 10 in the hybrid protein and mode 9 in trypsin are shown in Figure 6D. Mode 3 is similar in all of these proteins. Modes shown in Figure 6D are also similar in the loop region (190-194, 221-224). But in the region of residue 100-130, mode of chymotrypsin and the hybrid protein are similar. In the region of residue 170-180, mode of trypsin and the hybrid protein are similar. Although there are similarity and differences, single mode can not explain the correlation change of residue pairs that Figure 5 and Figure 6B have shown. Several modes work together to change the relationship of residue pairs.

4. Correlation plot

To get a detailed and more direct picture of the residue correlations, we "plot" the correlation directly onto the 3D structure. We use lines between two residues to illustrate the correlation between them (Fig. 7). Only large correlations (greater than 0.6) are shown with lines. We also omit the correlations if the distance between two residues is shorter than 7Å to emphasize the long range correlations. We note that residues 190-193 in chymotrypsin have strong correlation with residues 142-146 and residue 16 ("Loop D" region in Fig. 7A). In trypsin the correlation between 190-193 and the Loop D region is not as strong, and L1, L2 loops have certain correlations with the Loop D region (Fig. 7B). When the L1 and L2 loops in trypsin are changed into chymotrypsin loops, we find that the two loops no longer correlate strongly with the Loop D region. More importantly, the connections between the pocket residues

190-193 and the Loop D region become stronger, although these residues are intact in the virtual mutation (Fig. 7C). This means that loop substitution changes the dynamic correlations between residues 190-193 and residues on the Loop D region. This may have functional implications. Residue 192 is a residue in the S1 binding pocket, and it is important for inhibitor recognition in trypsin and chymotrypsin(63). On the Loop D there is an important residue Leu143 in chymotrypsin and Tyr151 in trypsin which are supposed to bind the P'2 residue of the substrate(37). Experiments show that in chymotrypsin S'2 site helps the reaction better than that in trypsin. S'2 is just on the Loop D and 190-193 is part of the S1 binding pocket. We know from the former analysis that chymotrypsin Loop D has stronger correlations with S1 pocket residue 190-193. This correlation will help to transfer the binding effect to the S1 site. In trypsin the correlation is weaker. This is consistent with the experiment (37). In this region, our analysis shows that the S1 binding pocket moves coherently with the residue contacting the P'2 site, similar to what we showed before: the S1 binding pocket moves coherently with the residue in the S3 site.

5. Conservation analysis

We extract 13 complete sequences of chymotrypsin and 64 sequences of trypsin from the ExPASy database (64). The sequence alignment was done using CLUSTAL_X(65) and the results are summarized in Table 1. The two loops are shown in black rectangles in Figure 1B. We notice that in both enzymes the length of Loop 1 is not conserved and the length of Loop 2 is conserved. In trypsin, L1 ranges 4-7 residues in length and L2 is 5 residues in length. In chymotrypsin, L1 ranges 4-5 residues and L2 is 4 residues in length.  The conservation of the length of L2 within chymotrypsin and trypsin may be important to the enzymes' selectivity.  Previous experiments support this idea.  In the experiment converting trypsin to chymotrypsin, trypsin with S1+L2 exchange is more active than S1+L1 mutant(1).  This means that L2 plays more important role than L1.  Compared with L1, L2 is shorter in most cases and not so extended, especially in chymotrypsin.  L2 links with the lid of S1 pocket, which is also a flexible component of the protein.  Thus transforming the motion of L2 to the S1 pocket is easier than that of L1.  If we calculate the correlation between S1 binding pocket and L1, L2 loops, we find the average correlation of L2-S1 is slightly stronger than L1-S1 correlation (about 0.005 stronger).

6. Dynamic property of loops and the substrate specificity of enzyme reaction

Correlation analysis shows that the motions of the two loops and the substrate-binding pocket are highly correlated.  The correlation between L1 and L2 in trypsin is mainly controlled by two major modes, whereas in chymotrypsin there are 5 major modes.  Loop motion of L1-L2 affects the dynamical relationship of S1 and Loop D.  The lengths of L1, L2 show very different conservations, which may be one of the reasons that L1 and L2 have different effects on enzyme specificity.  When trypsin was mutated at S1, L1 and L2 sites to those of chymotrypsin, the hybrid protein shows chymotrypsin-like loop correlations.  All the evidence implies that the dynamic property of the two loops play a critical role in making trypsin and chymotrypsin different.  This is in well accordance with the experiment(1) that shows that loop regions help to decide the specificity of chymotrypsin and trypsin. Miller and Agard(28) also reached the conclusion from a normal mode analysis that dynamics can be the determinant of substrate specificity in α–lytic protease. They found that the specificity of α–lytic protease correlates with the movement of the binding pocket. Molecular dynamics simulations also revealed the importance of the L1 and L2 loops in chymotrypsin catalysis: Wroblowski et al. showed that in both the activation and the deactivation of α-chymotrypsin, the targeted molecular dynamic path starts with a movement of loop 2, pulling on loop 1 (66). Both molecular dynamics simulation and modified GNM model have revealed that sites that are spatially distant from active sites can have a strong mechanical influence on the structural modulation of the substrate binding regions in HIV-1 protease (48).

[CONCLUSIONS]

We have studied the dynamical properties of trypsin and chymotrypsin and their relationship with enzyme specificity by using the Gaussian Network Model. A clustering method is introduced to analyze the correlations of the residues' motion. The two loops in trypsin and chymotrypsin were shown to have different dynamic properties which affect the correlations between other key sites in the two enzymes. When the two loops in trypsin were changed into chymotrypsin loops, the hybrid protein shows chymotrypsin-like cooperativity. Our results suggest that chymotrypsin like motions are important to the specificity of chymotrypsin. Changing the trypsin loops into chymotrypsin loops alters the motion style, and hence the specificity.

**Supplementary material**. The coordinates of the hybrid protein with the reconstructed loops can be found in the supplementary material.

**Acknowledgement**. This work was supported by the State Key Program of Basic Research of China (2003CB715900), the National Natural Science Foundation of China (90103029, 20173001, 20228306, 90403001, 30490240), the High-Tech Program of China, and the US National Science Foundation (DMR 0313129).

Table 1. Sequence length conservation of Loop 1 and Loop 2 in trypsin and chymotrypsin*.

|        | Protein \ loop length | 4    | 5    | 6   | 7   |
|--------|------------------------|------|------|-----|-----|
| Loop 1 | Trypsin                | 23%  | 0    | 66% | 11% |
|        | Chymotrypsin           | 54%  | 46%  | 0   | 0   |
|        | Protein \ loop length  | 4    | 5    | 6   | 7   |
| Loop 2 | Trypsin                | 0    | 100% | 0   | 0   |
|        | Chymotrypsin           | 100% | 0    | 0   | 0   |

*: 13 complete sequences of chymotrypsin and 64 sequences of trypsin from the ExPASy(64) database were used in the sequence alignment using CLUSTAL_X(65).

**Figure Legends**

**Figure 1. Superposition of trypsin and chymotrypsin.**

(A) The two enzymes have very similar tertiary structure. Trypsin is shown in green ribbon and chymotrypsin in blue. Active site residues of trypsin are shown in Ball& Stick. Loops of trypsin are shown in magenta; loops of chymotrypsin are shown in pale green. S1 binding pocket is shown in red. This figure is drawn using MOLMOL(67). (B) Sequence alignment of trypsin and chymotrypsin around the L1, L2 loop regions. Black shade indicates loops, gray shade indicates substrate binding pocket. Lowercase letters represent residues mutated in the experiments.

**Figure 2. Correlation map of chymotrypsin.**

Values of correlation between two residues range from –1 to 1. Blue means negative correlation and red means positive correlation, as shown in the color bar on the right. Both X-axis and Y-axis of this map are chymotrypsin residue indices. The two rectangles indicate the relative position of two beta barrels in the protein.

**Figure 3. Clustering analysis of chymotrypsin.**

(A) The mean square fluctuation of each mode. Note the value does not change much after mode 40, so we have used the first 40 modes in the calculation of correlations. (B) The tree of correlations of chymotrypsin. Residues form clusters and we draw a line to define these clusters for the plot in (Fig. 3C). (C) Different clusters are painted with different colors on the chymotrypsin structure. The colors are chosen arbitrarily.

**Figure 4. Local correlation trees of chymotrypsin, trypsin and the hybrid protein.**

Total length of horizontal lines between two residues is related to the correlation coefficient. The shorter the length, the stronger the two residues are correlated. (A)

The local correlation tree of chymotrypsin around the loop regions.   Residues on the two loops (shown with ●) cluster together with some of the residues in the S1 pocket (shown with ◄).   (B) The local tree of trypsin.   (C) The local tree of the hybrid protein.   In all these figures, many residues in the S1 binding pocket cluster with L1-L2 loops.   Figure 4C and Figure 4A are similar in that the correlations between residues 217-219 and L1-L2 loops are stronger in chymotrypsin and the hybrid protein than in trypsin.   Residues shown in lowercase letters are those mutated in experiment(1).   Figures are dawn by using TreeExplorer. (http://evolgen.biol.metro-u.ac.jp/TE/TE_man.html)

**Figure 5.   Comparison of pair-wise correlations among residues important for activity.**

This figure shows S value of some important residue pairs.   X-axis entries represent different residue pairs; corresponding Y-axis entry is the S value.   Most correlations of the hybrid protein are trypsin like but some correlations between key residues become chymotrypsin like.

**Figure 6.   Effect of selected modes on protein motion.**

(A) Contribution of the top modes to the loop region correlation.   X-axis is mode number, up to 40.   Larger numbered modes are not shown because they show little effect on the loop correlation.   Y-axis is the normalized ratio of the contribution.   (B) Fluctuations of residues calculated with the most important modes to the loop motion.   Modes 3 and 9 were used for trypsin.   Modes 3, 4, 5, 6 and 11 were used for chymotrypsin.   Modes 3, 4, 5, 9 and 10 were used for the hybrid protein.   (C) Mode 3 of the three proteins.   (D) Mode 11 in chymotrypsin, mode 10 in the hybrid protein and mode 9 in trypsin.

**Figure 7.   Correlations near the loop region.**

Correlations between two residues with an absolute value bigger than 0.7 are shown in lines. Correlations between 190-193 and Loop D are shown in red. (A) Chymotrypsin. (B) Trypsin. (C) The hybrid protein. In chymotrypsin and the hybrid protein, correlations shown in black are stronger than that in trypsin.

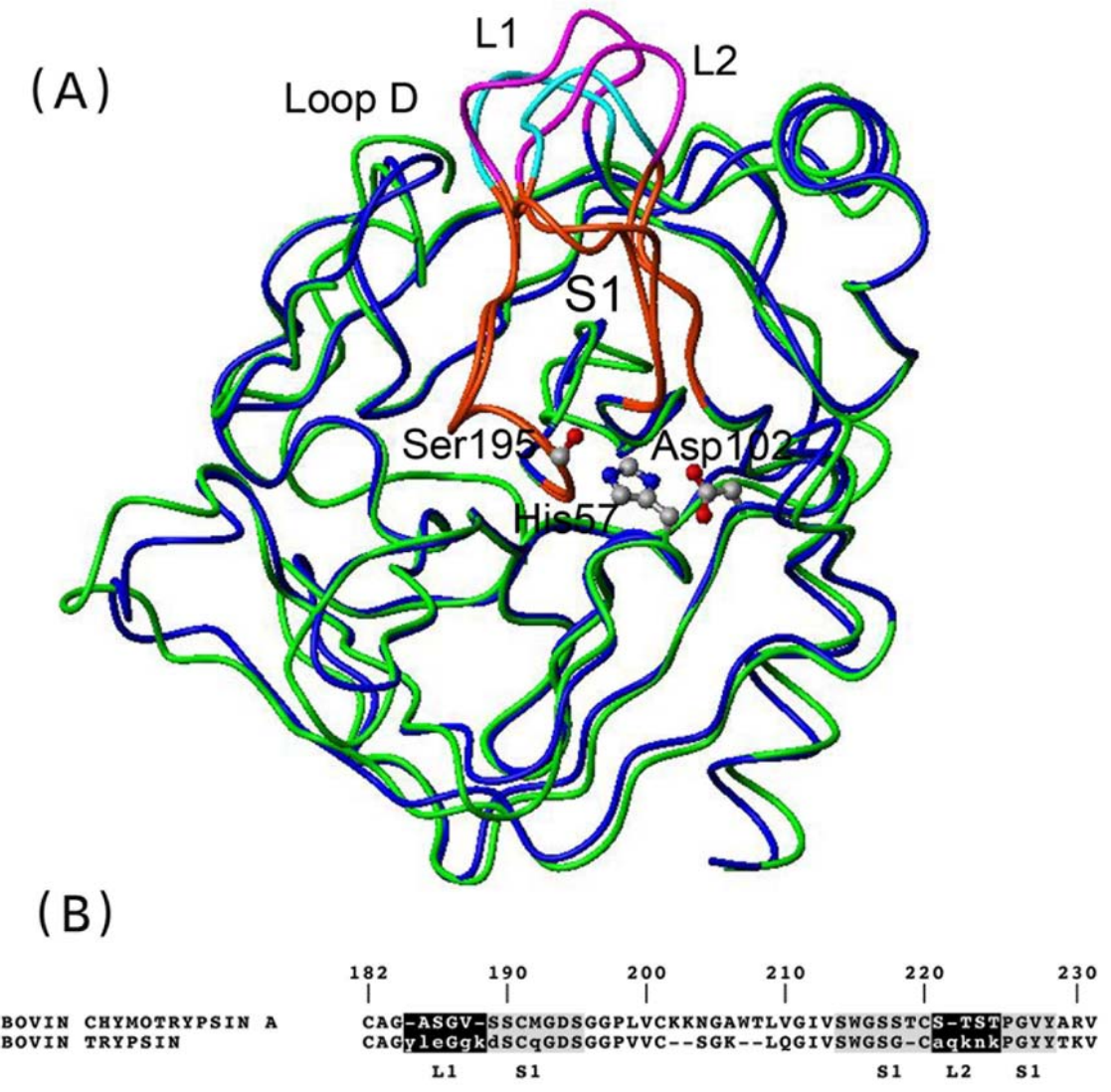

Figure 1

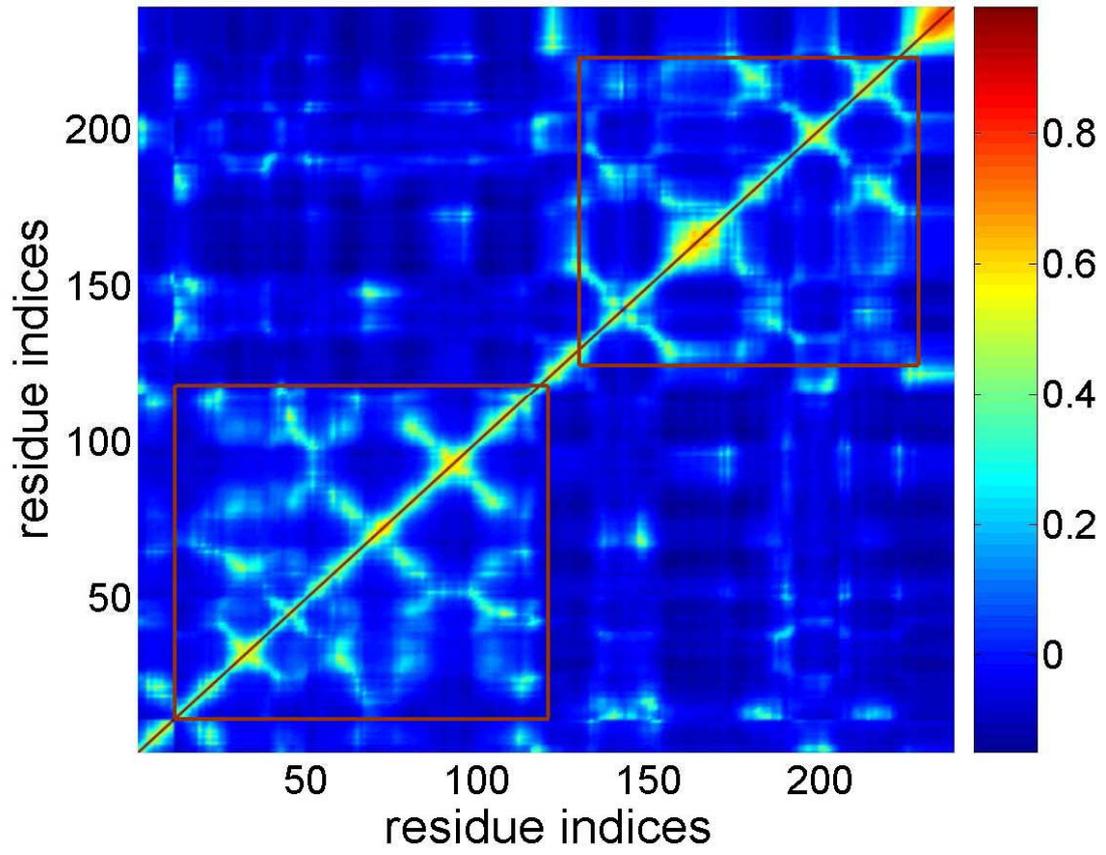

Figure 2

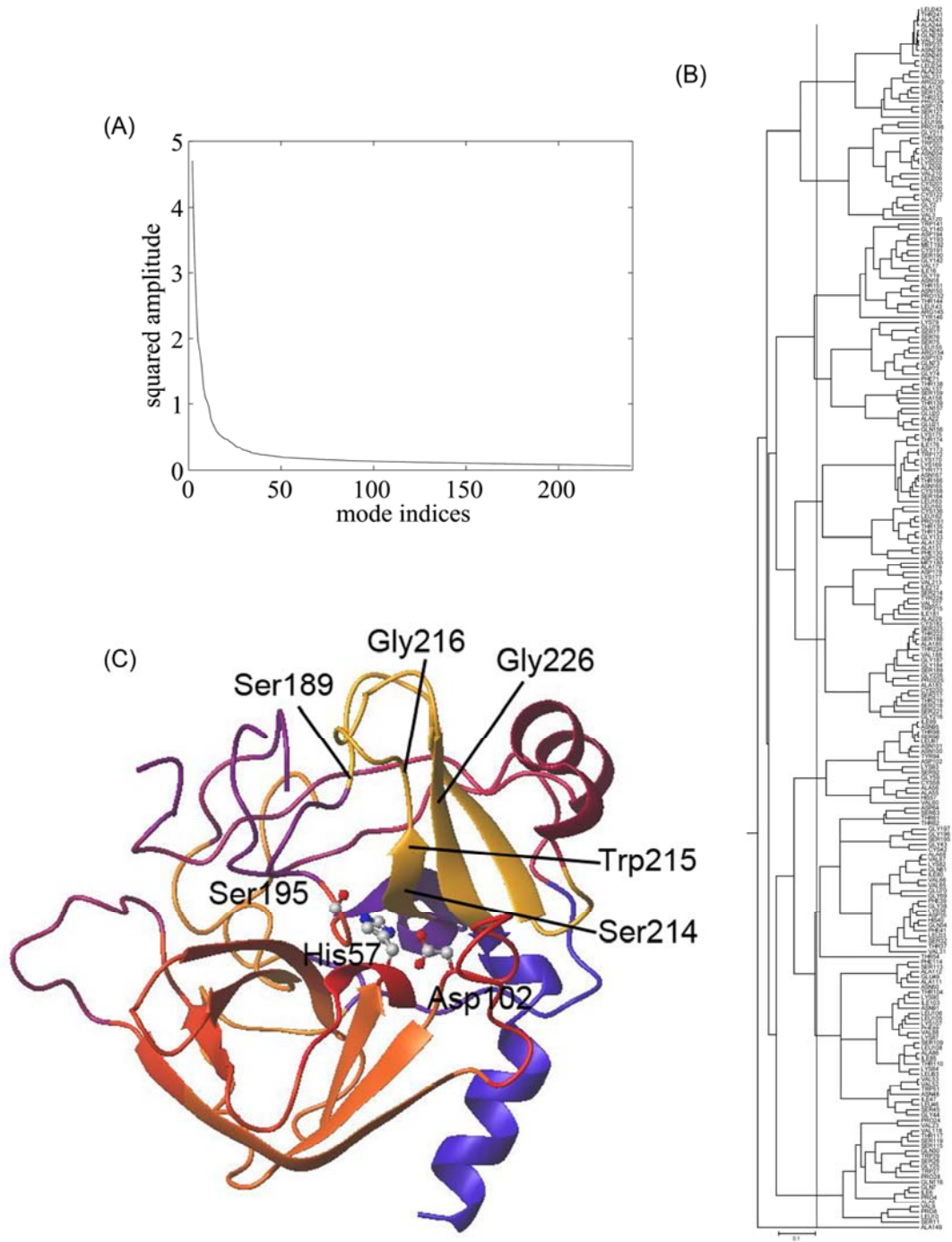

Figure 3

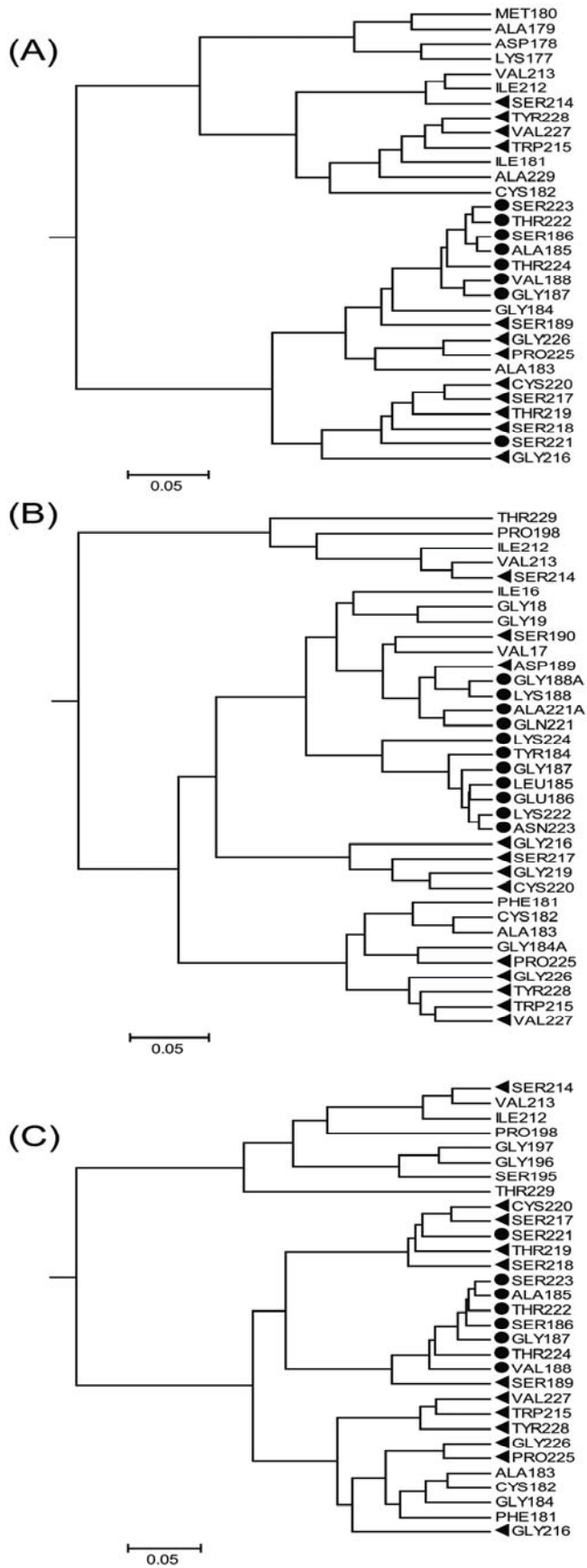

Figure 4

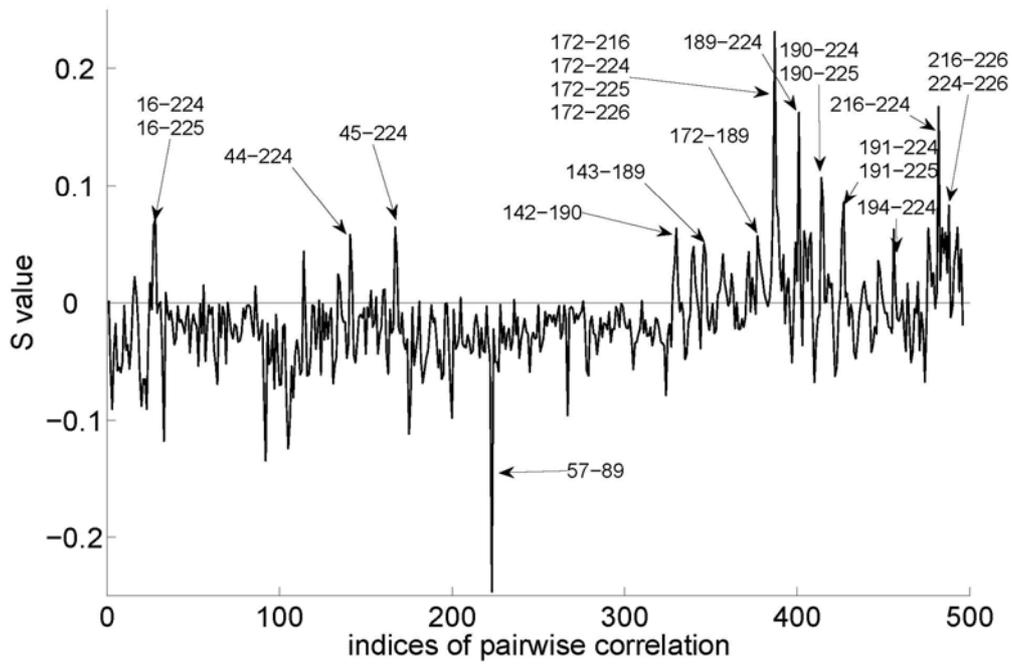

Figure 5

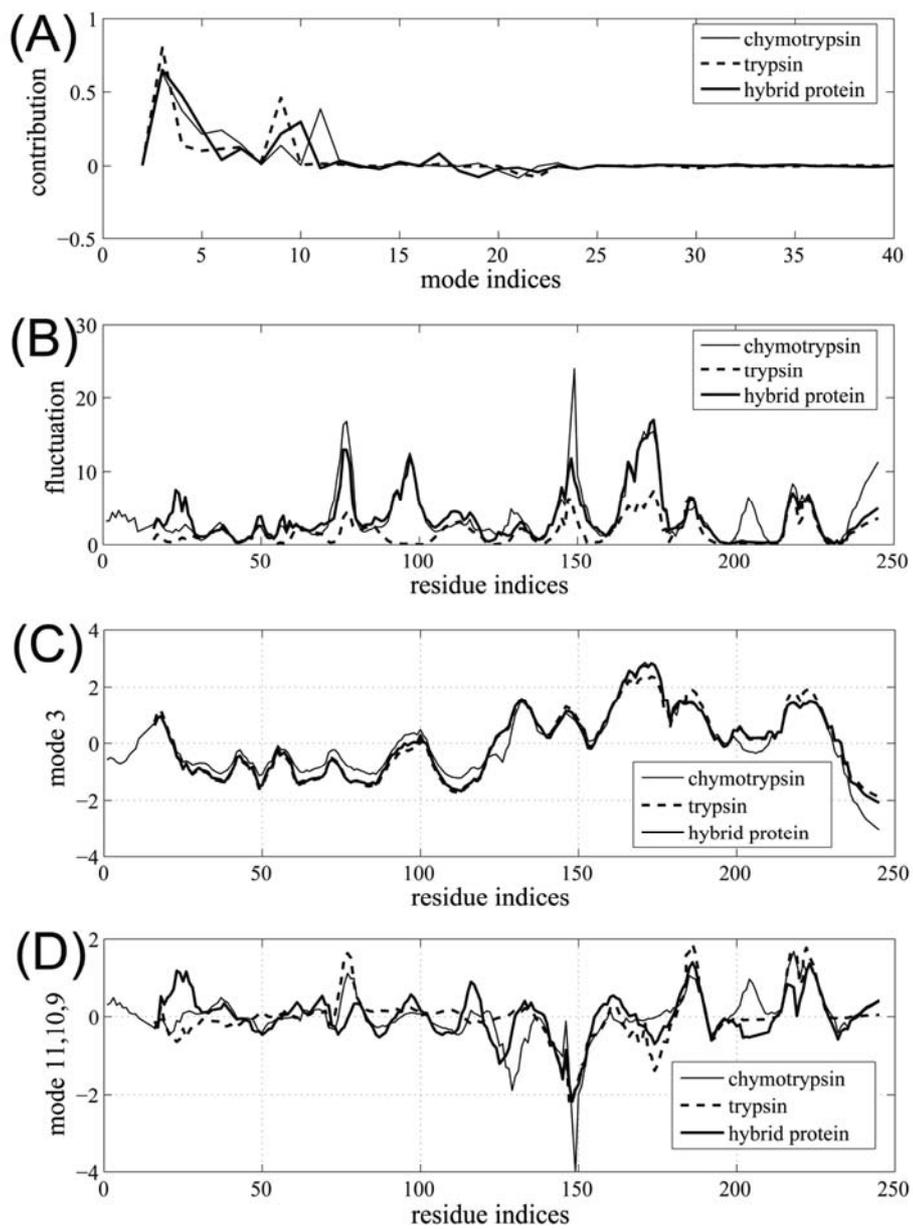

Figure 6

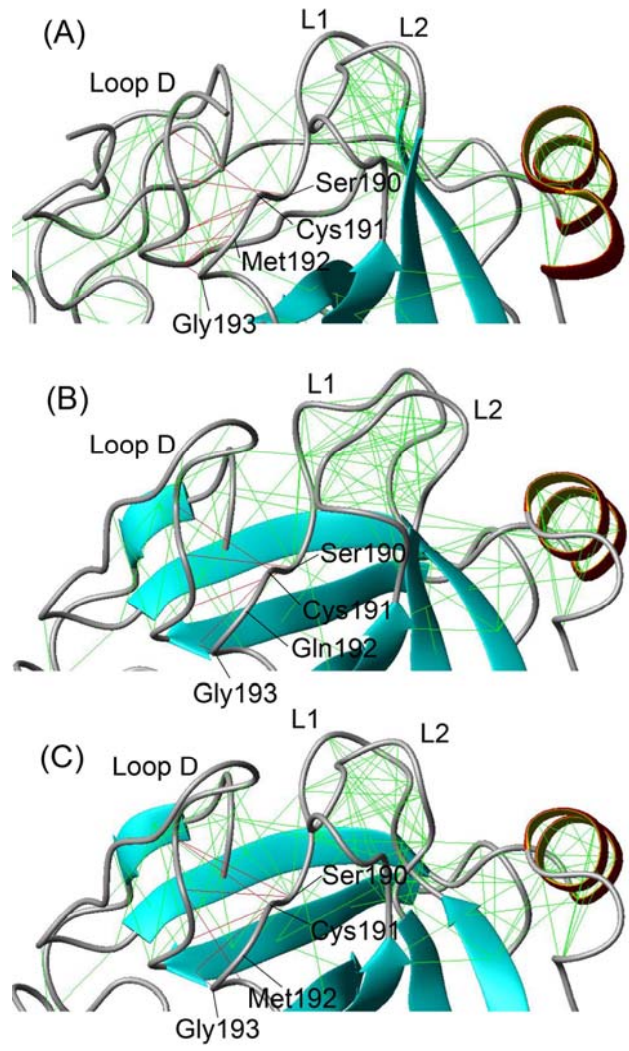

Figure 7